\documentclass[twoside]{article}
\usepackage{fleqn,espcrc2}

\newcommand{\p}[1]{(\ref{#1})}

\newcommand{\AmS}{{\protect\the\textfont2
  A\kern-.1667em\lower.5ex\hbox{M}\kern-.125emS}}

\title{Nonlinear Realizations
of Superconformal Groups and Spinning Particles}

\author{A. Pashnev\address{Bogoliubov Laboratory of 
       Theoretical Physics, JINR, \\
       Dubna, 141980, Russia}%
        \thanks{pashnev@thsun1.jinr.ru}%
        \thanks{Talk given  at the D.V. Volkov
        Memorial Conference ``Supersymmetry and Quantum Field Theory'',
        July 25-30, 2000, Kharkov}
        }

\begin{document}

\begin{abstract}
The method of nonlinear realizations is applied for the
conformally invariant description of the spinning particles in
terms of geometrical quantities of the parameter spaces of the one
dimensional $N$ - extended superconformal groups. We develop the
superspace approach to the cases of spin $0,\;\frac{1}{2},\;1$
particles and describe the alternative component approach in the
application to the spin-$\frac{1}{2}$ particle.
\vspace{1pc}
\end{abstract}

\maketitle

\section{INTRODUCTION}

The conformally invariant description of the relativistic
(spinning) particle treats on the same footings the coordinates of
the particle and (super)einbein, needed for the local
reparametrization invariance of the action \cite{M,S}. The
space-time coordinates in this approach are not fundamental ones.
They are ratios of some more basic variables, each transforming as
the one-dimensional vielbein. In total, for the description of the
particle in $D$ dimensions, one need $D+2$ such variables
(einbeins) entering the starting action identically.

As was shown in \cite{P,P1,P2}, these einbeins (and their
superpartners in the case of $N=1$ spinning particle) can be
naturally described as dilatons (with superpartners),
parametrizing $D+2$ different elements of the one-dimensional
superconformal group. In this paper we generalize the
consideration of \cite{P2} to the case of $N=2$ spinning particle
making use of the superspace approach. We also point out the
problems arising when attempts to generalize the approach to the
higher $N$ extensions of spinning particle
\cite{GT,HPPT}.
As a possible way out we describe the modified component approach
and apply it to the revision of the $N=1$ spinning particle.

The structure of the paper is the following. In Sect.2 we shortly
describe the nonlinear realizations method in general and mention
its peculiarities for the groups including the space - time
translations. As a simple application of the method we describe in
Sect.2.2 and Sect.2.3 the spinless
 and spin-$\frac{1}{2}$ particles in the superspace approach. We develop
in Sect.3  the superspace approach to the case of spin-$1$ $N=2$
spinning particle. In Sect.4 we analyze the possibility to
generalize the method to higher spins and describe the alternative
component approach in the application to the spin-$\frac{1}{2}$
particle. The algebra of
$N$-extended superconformal group is described in the Appendix.

\section{RELATIVISTIC MASSLESS PARTICLE AND $N=1$ SPINNING PARTICLE}

\subsection{The nonlinear realizations method}

In this method the representation space of the group coincides
with the whole space of the group parameters or with its subspace,
which describes the coset space of the group over some of its
subgroup. Let us consider the general group element and its
transformation under the left multiplication by some fixed element
$g$
\begin{equation}\label{G}
  G(a_i) \rightarrow G'=g\cdot G(a_i)=G(a'_i).
\end{equation}
So, the parameters $a_i$ of the group space realize the
representation of the group with the transformation law:
$a'_i =a'_i(a_k, g)$. If the group
admits the parametrization in the form
\begin{equation}\label{KH}
  G=K\cdot H = K(k_m)\cdot H(h_s),
\end{equation}
where $H$ is some subgroup of the group $G$ and $K$ parametrizes
the corresponding coset $K=G/H$, the  group action can be realized
not only on the whole space of parameters $\{k_m,h_s\}$, but on
the coset space
$\{k_m\}$ as well. The transformation law in this case is
\begin{equation}\label{TL}
  k'_m =k'_m(k_n, g),\quad h'_s =h'_s(k_m, h_t, g).
\end{equation}

 One can generalize the approach and consistently consider
simultaneously more than one group elements carrying the external
index ${\cal I}$
\begin{equation}\label{general}
G_{\cal I}=K\cdot H_{\cal I}
 \end{equation}
  with the same
coset element $K$ and different elements $ H_{\cal I}$ of the
subgroup $H$. This property (the equality of the coset elements
for all
$G_{\cal I}$ ) is invariant with respect to the left
multiplication
\begin{equation}\label{generaltransform}
G_{\cal I} \rightarrow G'_{\cal I}=g\cdot G_{\cal I}
\end{equation}
with any group element $g$.

The differential invariant Cartan's $\Omega$ - forms can be
constructed for each of these group elements
\begin{equation}\label{OI}
\Omega_{\cal I}=G_{\cal I}^{-1}dG_{\cal I}.
\end{equation}
Moreover, the following group elements (strictly speaking they
belong to the subgroup H)
\begin{equation}\label{IJ}
G_{\cal IJ}=G_{\cal I}^{-1}G_{\cal J}=H_{\cal I}^{-1}H_{\cal J}
\end{equation}
 are also invariant with respect to the left
multiplication \p{generaltransform}. This fact gives the
additional possibilities for construction of invariants of the
group. Let us underline that in contrast to the $\Omega$ - forms
\p{OI}, which belong to the algebra, new invariants \p{IJ} belong
to the group itself.

In what follows we will consider as starting groups the
superconformal infinitedimensional groups of the (super)spaces
with one bosonic and $N$ grassmann coordinates $Z^M =(\tau,
\theta_a)$. The corresponding (super)Virasoro algebras contain
among their generators the translation $P=L_{-1}$ and $N$
supercharges $G^a_{-1/2}$. In this case it is convenient to
parametrize the group element in the form
\begin{equation}\label{TrOut}
 G=e^{i\tau L_{-1}} \cdot e^{\theta_a G^a_{-1/2}}\cdot{\tilde G}
\end{equation}
and after that consider all parameters in ${\tilde G}$ as
functions in the superspace $(\tau, \theta_a)$. The coordinates
$(\tau, \theta_a)$ transform indeed as they should transform. Such
consideration automatically will lead to superfield constructions.
However, as we will see later, in some cases, especially when the
number of supersymmetries
$N$ growths, such superfield approach leads to difficulties.
Instead, one can consider already $\theta_a$ as the functions
$\theta_a(\tau)$ of a single bosonic coordinate $\tau$.
This situation corresponds to the phase of the spontaneously
broken supersymmetry and functions $\theta_a(\tau)$ play the role
of the corresponding Goldstone fields. Such component approach is
alternative to the superfield one and it, possibly, will help to
overcome the mentioned difficulties.

\subsection{Virasoro algebra and massless particle}

One dimensional diffeomorphisms algebra is the simplest example of
the $N$ - extended superconformal algebras described in the
Appendix. It coincide with the Virasoro (centerless) algebra
\begin{equation}\label{VA}
  \left[L_m,L_n\right]=-{\rm i}(m-n)L_{m+n}.
\end{equation}
If one limit themselves to the positive part of this algebra,
which generate in the one dimensional space the transformations
which are regular at the origin, the most natural is the following
parametrization of the group element
\begin{equation}\label{coset}
G=\!e^{i\tau L_{-1}} e^{iU^{(1)}L_1} e^{iU^{(2)}L_2}
e^{iU^{(3)}L_3}\ldots\! e^{i{U^{(0)}}L_0},
\end{equation}
in which all multipliers with the exception of $e^{i{U}L_0},\;
U\equiv U^{(0)}$, are ordered by the dimensionality $dim\; L_n=n$
of the correspondent generators.

Such structure of the group element simplifies the evaluation of
the variations $\delta {U^{(m)}}$ under the infinitesimal left
action
\begin{equation}\label{left}
G'=(1+i\epsilon)G,
\end{equation}
where $\epsilon =
\sum_{m=0}^\infty \epsilon^{(m)} L_{m-1}
$ belongs to the
algebra of the diffeomorphisms group. In particular
$\delta \tau= \sum_{m=0}^\infty \epsilon^{(m)}\tau^m \equiv \epsilon(\tau)$
All other
$U^{(n)}$ transforms through $\tau$ and $U^{(m)}, \; m<n$.

At this stage it is natural to consider all parameters
$U^{(n)}$ as the fields $U^{(n)}(\tau)$ in one dimensional space 
parametrized by the coordinate $\tau$. The field $U^{(0)}(\tau)$ 
transforms as a one dimensional dilaton. Simultaneously $U^{(1)}(\tau)$
transforms as one dimensional Cristoffel symbol.

Having in mind that the generators $L_0$ and $L_1$ form the
subalgebra, one can consider more than one group elements (${\cal
I}=0,1\ldots,D+1$)
\begin{equation}\label{ga}
G_{\cal I}=\!e^{i\tau L_{-1}}
 e^{iU^{(2)}L_2}
e^{iU^{(3)}L_3}\ldots\! e^{iU_{\cal I}^{(1)}L_1} e^{iU_{\cal
I}L_0},
\end{equation}
which have identical values of parameters
 $\tau$ and $U^{(m)}(\tau),\quad m\geq 2$, and
differ in the values of the parameters $U_{\cal I}^{(1)}$ and
$U_{\cal I}^{(0)}\equiv U_{\cal I}$.
This property is valid when all of these group elements are
transformed with the same infinitesimal transformation parameter
$\epsilon$ in \p{left}.

Consider the Cartan's differential form for each value of the
index
$\cal I$
\begin{equation}
\Omega_{\cal I}=G_{\cal I}^{-1}dG_{\cal I}=i\Omega_{\cal I}^{(-1)}L_{-1}+
i\Omega_{\cal I}^{(0)}L_0+\ldots.
\end{equation}
All its components $(\Omega_{\cal I}^{(-1)},\;\Omega_{\cal
I}^{(0)},\;
\Omega_{\cal I}^{(1)},\ldots)$
are invariant with respect to the left transformation
\p{generaltransform} (or \p{left}).
The explicit expressions for some components of the $\Omega$ -form
are:
\begin{eqnarray}\label{a}
\Omega_{\cal I}^{(-1)}
\!\!\!\!&=&\!\!\!\!
e^{-U_{\cal I}}d\tau,\\  \label{ab}
\Omega_{\cal I}^{(0)}
\!\!\!\!&=&\!\!\!\!
d U_{\cal I}-2d\tau U_{\cal I}^{(1)},\\ \label{abc}
\Omega_{\cal I}^{(1)}
\!\!\!\!&=&\!\!\!\!
(dU_{\cal I}^{(1)}+d\tau (U_{\cal I}^{(1)})^2-3d\tau
U^{(2)})e^{U_{\cal I}}.
\end{eqnarray}
The first of these forms is the differential one-form einbein. The
covariant derivatives (carrying the external index
${\cal I}$) calculated with its help are
\begin{equation}
D_{\tau\cal I}=e^{U_{\cal I}}\frac{d}{d\tau}.
\end{equation}

The most interesting is the form $\Omega_{\cal I}^{(1)}$. Using it
one can write the following invariant expression for the action
\begin{eqnarray} \label{act}
S\!\!\!\!&=&\!\!\!\!
-\frac{1}{2}\Sigma_{\cal I}\int \Omega_{\cal I}^{(1)}=\\
\!\!\!\!&=&\!\!\!\!
-\frac{1}{2}\Sigma_{\cal I}\int d \tau e^{U_{\cal I}}
(\dot{U}_{\cal I}^{(1)}+ (U_{\cal I}^{(1)})^2-3U^{(2)}),\nonumber
\end{eqnarray}
where $\Sigma_{\cal I}$ is the signature of $D+2$ - dimensional
space-time (with two times)
\begin{equation}\label{sigma}
\Sigma_{\cal I}=(-\underbrace{++\ldots ++}_{\mbox{$D$}}-)
\end{equation}
and summation over external index ${\cal I}$ is understood.

After the elimination of $U_{\cal I}^{(1)}$ with the help of its
equation of motion the action \p{act} in terms of new variables
\begin{equation}
x_{\cal I}=e^{U_{\cal I}(\tau)/2},\;\;
\lambda=-3U^{(2)}(\tau)
\end{equation}
has the form
\begin{equation}               \label{1S}
S=\int{\rm d} \tau(\frac{1}{2}\dot{x}^2_{\cal I}-
\frac{1}{2}\lambda x^2_{\cal I}).
\end{equation}
The relation of the action (\ref{1S}) with the usual\\ $D$ -
dimensional action is established  by solving the equation of
motion for the Lagrange multiplier $\lambda$ \cite{M}
\begin{equation}\label{x^2}
  x^2_{\cal I}=0.
\end{equation}
Note the triviality of  dynamics implied by this equation in the
absence of the external index ${\cal I}$ as well as in the case of
positive definite signature $\Sigma_{\cal I}$.

In terms of new variables
\begin{eqnarray}\label{4S}
\tilde{x_i}\!\!\!\!&=&\!\!\!\!\frac{x_i}{x_+},\;\; \;e=\frac{1}{x_+^2},
 \;\;\;
(x_-=-\frac{x^ix_i}{2x_+}),\\
\nonumber
x_\pm\!\!\!\!&=&\!\!\!\! \frac{1}{\sqrt{2}}(x_D\pm x_{D+1}),
(i=0,1,...D-1),
\end{eqnarray}
the Lagrangian  in (\ref{1S}) becomes the standard one
\begin{equation}
L=\frac{1}{2}\;\frac{\dot{\tilde{x}}^2}{e}.
\end{equation}
The expressions \p{4S} for coordinates $\tilde{x_i}$ show that
they are indeed scalars with respect to the transformations of the
one dimensional diffeomorphisms group. At the same time $e(\tau)$
transforms correctly as an einbein. All this is a result of the
transition from $D+2$ dimensional to $D$ dimensional consideration
which is implied by the constraint $ x^2_{\cal I}=0$.

\subsection{$N=1$ spinning particle in the superspace approach}
To generalize the approach  on the spinning particles we  consider
the $N=1$ SCA which is the simplest of the algebras described in
the Appendix. Its  generators are placed on the first two lines of
the Picture 1 and have the following commutation relations in
addition to \p{VA}
\begin{eqnarray}
\left[L_m,G_s\right]&=&-{\rm i}(\frac{m}{2}-s)G_{m+s}\\
\{G_r,G_s\}&=&2L_{r+s}.
\end{eqnarray}
Following the considerations of the previous subsections we write
$D+2$ group elements as (${\cal I}=0,1\ldots,D+1$)
\begin{eqnarray}
G_{\cal I}\!\!\!\!&=&\!\!\!\!e^{{\rm i}\tau L_{-1}} \cdot
 e^{{\rm i}\theta G_{-1/2}}
 \cdot e^{{\rm i}\Theta^{(3/2)}G_{3/2}}
 \cdot e^{{\rm i}U^{(2)}L_2} \cdots \nonumber\\
\!\!\!\!&&\!\!\!\!
e^{{\rm i}\Theta_{\cal I}G_{1/2}}  \cdot e^{{\rm i}U_{\cal
I}^{(1)}L_1}
 \cdot e^{{\rm i}U_{\cal I}L_0}.
\label{GA}
\end{eqnarray}
Last three multipliers in this expression form the subgroup of the
whole $N=1$ superconformal group and they consistently can carry
external index ${\cal I}$. All parameters (Grassmann $\Theta$-s
and commuting $U$-s) are considered as superfunctions in the
$(1,1)$ superspace parametrized by $\tau$ and $\theta$. The
variations of these superspace coordinates under the left action
of infinitesimal superconformal transformation are given by the
general expressions
\p{InfT}.

To calculate the invariant differential $\Omega$- forms one should
take into account that Grassmann parity of differential of any
variable is opposite to its own Grassmann parity, i.e.
$d\tau$ is odd and $d\theta$ is even \cite{B}.
The general expression for Cartan's $\Omega$ - form is now
\begin{eqnarray}\nonumber
\Omega_{\cal I}\!\!\!\!&=&\!\!\!\!
G_{\cal I}^{-1}dG_{\cal I}={\rm i}\Omega_{\cal I}^{(-1)}L_{-1}+
{\rm i}\Omega_{\cal I}^{(-1/2)}G_{-1/2}+\\
\!\!\!\!&&\!\!\!\!
{\rm i}\Omega_{\cal I}^{(0)}L_0+
{\rm i}\Omega_{\cal I}^{(1/2)}G_{1/2}+{\rm i}\Omega_{\cal
A}^{(1)}L_1+\ldots.
\end{eqnarray}
Its two first components
\begin{eqnarray}
\Omega_{\cal I}^\tau
\!\!\!\!&\equiv&\!\!\!\!
\Omega_{\cal I}^{(-1)}
=({\rm d}\tau-{\rm i}{\rm d}\theta\theta)e^{-U_{\cal I}}=
{\rm d}x^M{E_M^\tau}_{\cal I} ,\\
\nonumber
\Omega_{\cal I}^\theta
\!\!\!\!&\equiv&\!\!\!\!
\Omega_{\cal I}^{(-1/2)}
=\{{\rm d}\theta-({\rm d}\tau-{\rm id}\theta\theta)\Theta_{\cal
I}\}e^{-U_{\cal
I}/2}\\
\!\!\!\!&=&\!\!\!\!
{\rm d}x^M{E_M^\theta}_{\cal I}
\end{eqnarray}
define supervielbeins ($x^1\equiv\tau, x^2\equiv\theta):$\vspace{0.3cm}\\
\begin{tabular}{c|ll|r}
  & $e^{-U_{\cal I}}$ & $-\Theta_{\cal I}\cdot e^{-U_{\cal I}/2}$&\\
${E_M^A}_{\cal I}=$&&&\\
  & $-{\rm i}e^{-U_{\cal I}}\cdot\theta$ &
$e^{-U_{\cal I}/2}(1-{\rm i}\Theta_{\cal I}\cdot\theta)$&\\
\end{tabular}\vspace{0.3cm}\\

The invariant integration measure is
\begin{equation}
{\rm d}V_{\cal I}={\rm d}\tau\underline{{\rm d}\theta}Ber({E_M^A}_{\cal I}),
\end{equation}
where $\underline{{\rm d}\theta}$ is the Berezin differential and
\begin{equation}
Ber({E_M^A}_{\cal I})=e^{-U_{\cal I}/2}.
\end{equation}
Note, that the integration measure, as well as the supervielbeins,
depend on the external index ${\cal I}$.

The action for $N=1$ spinning particle is constructed by using the
coefficient $\Gamma_{\cal I}$ in the expression of  the invariant
component $\Omega_{\cal I}^{(1)}$ in terms of the full system of
invariant differential forms $\Omega_{\cal I}^\tau$ and
$\Omega_{\cal I}^\theta$:
\begin{equation}
\Omega_{\cal I}^{(1)}=\Omega_{\cal I}^{\tau}Y_{\cal I}+
\Omega_{\cal I}^{\theta}\Gamma_{\cal I}.
\end{equation}
This odd coefficient, as well as the even one $Y_{\cal I}$, is
also invariant. The invariant action is \cite{P,P2}
\begin{eqnarray} \label{scact}
S\!\!\!\!&=&\!\!\!\!\frac{{\rm i}}{2}\Sigma_{\cal I}\int {\rm
d}V_{\cal I}
\Gamma_{\cal I}=\\
&&\!\!\!\!\!\!\!\!\frac{i}{2}\Sigma_{\cal I}
\int {\rm d}\tau {\rm d}\underline{\theta}
e^{U_{\cal I}} (D_\theta U^{(1)}_{\cal I}-{\rm
i}D_\theta\Theta_{\cal I}
 \Theta_{\cal I}+\nonumber\\
&&\!\!\!\!\!\!\!\!2{\rm i}\Theta_{\cal I}U^{(1)}_{\cal I}- 2{\rm
i}\Theta^{(3/2)}).\nonumber
\end{eqnarray}
After the solution of the equations of motion for auxiliary fields
$U^{(1)}_{\cal I}$,
and introduction of new variables
$X_{\cal I}=e^{U_{\cal I}/2}=
x_{\cal I}(\tau)+{\rm i}\theta\gamma_{\cal I}(\tau)$ the action
\p{scact} becomes
\begin{equation}\label{S1}
S=-\frac{{\rm i}}{2}\int {\rm d}\tau {\rm d}\underline{\theta}
(\dot{X_{\cal I}}D_\theta X_{\cal I} +2i\Theta^{3/2} X^2_{\cal
I}).
\end{equation}
The superfield $\Theta^{3/2}(\tau,\theta)=
\frac{1}{2}(\rho-\theta\lambda)$ in this action plays
the role of Lagrange multiplier leading to the constraint
$X^2_{\cal I}=0$ which is the supersymmetric generalization of the
constraint \p{x^2}. After the Berezin integration over the
$\theta$ the action \p{S1} coincides with the manifestly conformal
component action for the $N=1$ spinning particle \cite{M,S}
\begin{equation}               \label{1SC}
S=\frac{1}{2}\int{\rm d} \tau(\dot{x}^2_{\cal I}+
i\dot\gamma_{\cal I}\gamma_{\cal I}-
\lambda x^2_{\cal I}-2i\rho\gamma_{\cal I}x_{\cal I}).
\end{equation}

\section{$N=2$ SUPERCONFORMAL ALGEBRA AND SPIN-$1$ PARTICLE
IN THE SUPERSPACE APPROACH}

$N=2$ Superconformal Algebra (SCA) in complex notations
\begin{equation}\label{CN}
 G_s=G_s^1+iG_s^2,\quad T_m=-\frac{i}{2}T^{12}_m,
\end{equation}
(see Appendix) have the following form
\begin{eqnarray}
\left[L_m,L_n\right]&=&-i(m-n)L_{m+n}, \nonumber\\
\left[L_m,T_n\right]&=&imT_{m+n}\nonumber\\
\left[L_m,G_s\right]&=&-i(\frac{m}{2}-s)G_{m+s}, \nonumber\\
\left[L_m,{\bar G}_s\right]&=&-i(\frac{m}{2}-s){\bar G}_{m+s}\\
\left[T_m,G_s\right]&=&-\frac{i}{2}G_{m+s},  \nonumber\\
\left[T_m,{\bar G}_s\right]&=&-\frac{i}{2}{\bar G}_{m+s}\nonumber\\
\{G_r,{\bar G}_s\}&=&2L_{r+s}+2T_{r+s}.  \nonumber
\end{eqnarray}
Following the previous considerations we consider simultaneously
$D+2$ elements of $N=2$ Superconformal group $({\cal I}=0,1\ldots,D+1)$
\begin{eqnarray}\nonumber
G_{\cal I}\!\!\!\!&=&\!\!\!\!e^{i\tau L_{-1}} \cdot e^{\bar\theta
G_{-1/2}+
\theta {\bar G}_{-1/2}} \cdot e^{V^1T_1}
 \cdot \\
\!\!\!\!&&\!\!\!\! e^{\bar\Theta^{(3/2)}G_{3/2}+
\Theta^{(3/2)}{\bar G}_{3/2}}
 \cdot e^{iU^{(2)}L_2} \cdots \\
\!\!\!\!&&\!\!\!\!
e^{\bar\Theta_{\cal I}G_{1/2}+\Theta_{\cal I}{\bar G}_{1/2}}
\cdot e^{iU_{\cal I}^{(1)}L_1}
 \cdot e^{iU_{\cal I}L_0} \cdot e^{V_{\cal I}T_0}.\nonumber
\end{eqnarray}
Last four multipliers in this expression form the subgroup of the
whole $N=2$ superconformal group and they consistently can carry
external index ${\cal I}$.

The line of calculations is the same. Firstly we find the
expressions for Cartan's $\Omega$ - form components
\begin{eqnarray}\nonumber
\Omega_{\cal I}\!\!\!\!&=&\!\!\!\!
G_{\cal I}^{-1}dG_{\cal I}={\rm i}\Omega_{\cal I}^{(-1)}L_{-1}+\\
\!\!\!\!&&\!\!\!\!
\bar\Omega_{\cal I}^{(-1/2)}G_{-1/2}+
\Omega_{\cal I}^{(-1/2)}{\bar G}_{-1/2}+\\
\nonumber
\!\!\!\!&&\!\!\!\!
{\rm i}\Omega_{\cal I}^{(0)}L_0+\Omega_{\cal I}^{T}T_0+\\
\nonumber
\!\!\!\!&&\!\!\!\!
\bar\Omega_{\cal I}^{(1/2)}G_{1/2}+
\Omega_{\cal I}^{(1/2)}{\bar G}_{1/2}+\ldots.\nonumber
\end{eqnarray}
Its three first components
\begin{eqnarray}\label{N2V1}
\Omega_{\cal I}^\tau
\!\!\!\!&\equiv&\!\!\!\!
\Omega_{\cal I}^{(-1)}={\rm d}x^M{E_M^\tau}_{\cal I}=\\
\!\!\!\!&=&\!\!\!\!
({\rm d}\tau-{\rm i}{\rm d}\theta\bar\theta-{\rm i}{\rm
d}\bar\theta\theta)
e^{-U_{\cal I}},\nonumber\\
\Omega_{\cal I}^\theta
\!\!\!\!&\equiv&\!\!\!\!
\Omega_{\cal I}^{(-1/2)}
={\rm d}x^M{E_M^\theta}_{\cal I}\\
\!\!\!\!&=&\!\!\!\!
\{{\rm d}\theta-({\rm d}\tau-{\rm i}{\rm d}\theta\bar\theta-
{\rm i}{\rm d}\bar\theta\theta)\Theta_{\cal I}\}
e^{-U_{\cal I}/2-{\rm i} V_{\cal I}/2 },\nonumber\\\label{N2V3}
\Omega_{\cal I}^{\bar\theta}
\!\!\!\!&\equiv&\!\!\!\!
\bar\Omega_{\cal I}^{(-1/2)}
={\rm d}x^M{E_M^{\bar\theta}}_{\cal I}\\
\!\!\!\!&=&\!\!\!\!
\{{\rm d}\bar\theta-({\rm d}\tau-{\rm i}{\rm d}\theta\bar\theta-
{\rm i}{\rm d}\bar\theta\theta)\bar\Theta_{\cal I}\}
e^{-U_{\cal I}/2+{\rm i} V_{\cal I}/2 }\nonumber
\end{eqnarray}
define supervielbein ${E_M^A}_{\cal I}$ in the notations:
$x^1\equiv\tau, x^2\equiv\theta, x^3\equiv\bar\theta.$

The invariant integration measure is simply
\begin{equation}
{\rm d}V_{\cal I}={\rm d}\tau\;\underline{{\rm d}\theta}\;
\underline{{\rm d}\bar\theta},
\end{equation}
because $Ber({E_M^A}_{\cal I})=1$ for the case of $N=2$ SCA, as
one can calculate using the expressions \p{N2V1}-\p{N2V3}. For
construction of the action one need also the expressions
\begin{eqnarray}
\Omega_{\cal I}^{(1/2)}
\!\!\!\!&=&\!\!\!\!
\{
{\rm d}\theta_{\cal I}-
({\rm d}\theta- ({\rm d}\tau-{\rm i}{\rm d}\theta\bar\theta-
{\rm i}{\rm d}\bar\theta\theta)
\Theta_{\cal I})
U^1_{\cal I}\nonumber\\\label{O1}
\!\!\!\!&&\!\!\!\!  -i {\rm d}\theta {\bar\theta}_{\cal I}
\theta_{\cal I}-\frac{i}{2}{\rm d}\theta V^1
\}
e^{U_{\cal I}/2-{\rm i} V_{\cal I}/2 },\\
\bar\Omega_{\cal I}^{(1/2)}
\!\!\!\!&=&\!\!\!\!
\{
{\rm d}\bar\theta_{\cal I}-
({\rm d}\bar\theta- ({\rm d}\tau-{\rm i}{\rm d}\theta\bar\theta-
{\rm i}{\rm d}\bar\theta\theta)
\bar\Theta_{\cal I})
U^1_{\cal I}\nonumber\\\label{O2}
\!\!\!\!&&\!\!\!\!  +i {\rm d}\bar\theta {\bar\theta}_{\cal I}
\theta_{\cal I}+\frac{i}{2}{\rm d}\bar\theta V^1
\}
e^{U_{\cal I}/2-{\rm i} V_{\cal I}/2 },
\end{eqnarray}
Their expansion in terms of fundamental forms \p{N2V1}-\p{N2V3}
contains invariant coefficients
\begin{eqnarray}
\Gamma
\!\!\!\!&=&\!\!\!\!
\{
D\theta_{\cal I}- U^1_{\cal I} -i  {\bar\theta}_{\cal I}
\theta_{\cal I}-\frac{i}{2} V^1
\}
e^{U_{\cal I}},\\
\bar\Gamma
\!\!\!\!&=&\!\!\!\!
\{
\bar{D}\bar\theta_{\cal I}-
U^1_{\cal I} +i  {\bar\theta}_{\cal I}
\theta_{\cal I}+\frac{i}{2} V^1
\}
e^{U_{\cal I}},
\end{eqnarray}
which can be used for the construction of the action
\begin{eqnarray}\label{AS}
  S_{N=2}\!\!\!\!&=&\!\!\!\!
\frac{i}{4}\int {\rm d}\tau\underline{{\rm d}\theta}
\underline{{\rm d}\bar\theta}(\Gamma-\bar\Gamma)=
\int \;{\rm d}V \;{\cal L},\\
\label{LS}
{\cal L}\!\!\!\!&=&\!\!\!\!
\frac{i}{4}\{D\theta_{\cal I}-{\bar D}\bar\theta_{\cal I}-
2i {\bar\theta}_{\cal I}
\theta_{\cal I}-i V^1\}
e^{U_{\cal I}}.
\end{eqnarray}
After the solution of the equations of motion for auxiliary fields
$\theta_{\cal I},\; \bar\theta_{\cal I}$ this lagrangian
becomes
\begin{equation}\label{S2}
{\cal L}=\frac{{1}}{2}DX_{\cal I}\bar{D} X_{\cal I} +
\frac{1}{4}V^1 X^2_{\cal I}).
\end{equation}
Here
$X_{\cal I}=e^{U_{\cal I}/2}=
x_{\cal I}(\tau)+{\rm i}\bar\theta\gamma_{\cal I}(\tau)+
{\rm i}\bar\theta\gamma_{\cal I}(\tau)+\bar\theta\theta F_{\cal I}(\tau)$
are the $N=2$ superfield coordinates and
$D=\partial/\partial\theta+i\bar\theta\partial/\partial\tau,\quad
{\bar D}=\partial/\partial\bar\theta+i\theta\partial/\partial\tau$
- the flat covariant derivatives. The $N=2$ superfield
$V^1(\tau,\theta, \bar\theta$ in this lagrangian plays the role of
Lagrange multiplier leading to the constraint
$X^2_{\cal I}=0$ which is the $N=2$ supersymmetric generalization of the
constraint \p{x^2} \cite{S}. The integration over the grassmann
coordinates, normalized as
$\int \underline{{\rm d}\theta} \underline{{\rm d}\bar\theta}\;
\bar\theta \theta=-1$ leads to the $D+2$ -  dimensional component 
lagrangian for $N=2$ spinning particle received in \cite{S}.

\section{$N=1$ SPINNING PARTICLE IN THE COMPONENT APPROACH}

The analysis of the possible generalization of described scheme on
higher $N$ spinning particles reveals the following obstacle. In
all considered examples the dimensionality of the action in the
units $dim \;\tau=+1, dim \;\theta_a=+1/2$ is $dim \; S=-1$. The
dimensionality of the integration volume for $N$ extended
supersymmetry is $dim \;{\rm d} V= 1-N/2$ whereas the
dimensionality of the components of Cartan's $\Omega$ - form is
nonpositive integers or halfintegers. It means that starting with
$N=4$ there do not exist the appropriate invariant coefficients
in the expansions of these $\Omega$ - form components in terms of
vielbeins which can be taken as the lagrangian (see the Picture
1).

One possible way out consists in the using as lagrangians the more
complicated (nonlinear) functions of these invariant coefficients.
The structure of these functions in the every case needs the
additional analysis. However, there exists the universal approach
to all cases of $N$ - extended supersymmetry. This is the so
called component approach, in which all parameters of the group
are the functions of a parameter $\tau$ (the proper time) only. In
some sense this approach is more economic, because it reduces the
number of the fields by $2^N$ times (the number of component
fields in a superfield). The only price, as was described in the
Sect.2, is the appearance of the grassmann Goldstone fields, which
corresponds to the $\tau$ dependent parameters at the
supertranslations generators. In all the cases the action can have
the form analogous to
\p{act}
\begin{equation}\label{GenAct}
  S=-\frac{1}{2}\Sigma_{\cal I}\int \Omega_{\cal I}^{(1)}
\end{equation}
where $\Omega_{\cal I}^{(1)}$ is the component of the $\Omega$ -
form corresponding to the generator $L_1$. So defined action by
construction is the supersymmetrization of the spinless particle
action. The only thing one should to do - is to redefine the
variables in such a way that the dependence of the action on the
grassmann Goldstone fields disappears. Below we illustrate this
approach for the simplest case of $N=1$ supersymmetry.

Once more consider the parametrization of the $D+2$ elements of
the $N=1$ SCA. This time we consider the spontaneously broken
realization of the supersymmetry transformation, i.e. the
corresponding parameter instead of being the Grassmann coordinate
of the superspace $(1,1)$ is now the Goldstone field
$\vartheta(\tau)$ which depends on the only bosonic coordinate
$\tau$
\begin{eqnarray}
G_{\cal I}\!\!\!\!&=&\!\!\!\!e^{{\rm i}\tau L_{-1}}
 e^{{\rm i}\vartheta(\tau) G_{-1/2}}
 e^{{\rm i}\Theta^{(3/2)}G_{3/2}}
 e^{{\rm i}U^{(2)}L_2} \cdots \nonumber\\
\!\!\!\!&&\!\!\!\!
e^{{\rm i}\Theta_{\cal I}G_{1/2}}  e^{{\rm i}U_{\cal I}^{(1)}L_1}
 e^{{\rm i}U_{\cal I}L_0}.
\label{GA1}
\end{eqnarray}
All other parameters are the functions of $\tau$ as well.

The explicit expression for the $\Omega_{\cal I}^{(1)}$ component
is the following
\begin{eqnarray}\label{GA2}
\Omega_{\cal I}^{(1)}
\!\!\!\!&=&\!\!\!\! \{dU_{\cal I}^{(1)}+d\tau
(U_{\cal I}^{(1)})^2-3d\tau
U^{(2)}-\\
\!\!\!\!&&\!\!\!\! i d \Theta_{\cal I}\Theta_{\cal I} +
2 i d \vartheta \Theta_{\cal I}U_{\cal I}^{(1)}- id
\vartheta\vartheta (U_{\cal I}^{(1)})^2 +\nonumber\\\nonumber
\!\!\!\!&&\!\!\!\! 3id \vartheta\vartheta U^{(2)}-
2i d \vartheta \Theta^{(3/2)}+\\
\!\!\!\!&&\!\!\!\!
2i d\tau \Theta_{\cal I}\Theta^{(3/2)}+ 2 d \vartheta\vartheta
\Theta_{\cal I}\Theta^{(3/2)}
\}e^{U_{\cal I}}.\nonumber
\end{eqnarray}
One can eliminate the auxiliary fields $U_{\cal I}^{(1)}$ and
introduce new fields (the dot denote the $\tau$-derivative)
\begin{eqnarray}\label{NF}
x_{\cal I}\!\!\!\!&=&\!\!\!\!e^{U_{\cal I}/2}(1-
i \vartheta \Theta_{\cal I}),\\
\gamma_{\cal I}\!\!\!\!&=&\!\!\!\!-e^{U_{\cal I}/2}((1-
i \dot\vartheta\vartheta)\Theta_{\cal I}+
\frac{1}{2}\dot{U}_{\cal I}e^{U_{\cal I}/2}\vartheta,\\
\lambda\!\!\!\!&=&\!\!\!\!\!\!-3U^{(2)}(1- i \dot\vartheta\vartheta)\!\!-
4i\dot\vartheta\Theta^{(3/2)}\!\!
-2i\vartheta\dot\Theta^{(3/2)},\\
\rho\!\!\!\!&=&\!\!\!\!-2\Theta^{(3/2)}
(1-3i\dot\vartheta\vartheta)-6\vartheta U^{(2)}.
\end{eqnarray}
In terms of these new variables the action \p{GenAct} coincides
with the manifestly conformal form of the action for $N=1$
spinning particle \p{1SC} \cite{M,S}.

\section{CONCLUSIONS}

All considered examples illustrate the close connection between
the physical systems and their symmetry groups, which consists in
the possibility to describe the system in terms of the parameters
of its symmetry group. One more well known example is the gravity
which can be described in terms of the metric tensor \cite{BO}
(see also \cite{IN} for supergravity) or vielbein \cite{P_0}
parametrizing the diffeomorphisms group of the space-time.

So, it would be interesting to apply the approach developed here
to the cases when the bosonic part of the (super)space is not one
dimensional. The simplest are the two dimensional spaces, which
correspond to the (super)strings. In addition, the method can be
applied to the nonlinearly realized  W-algebras which are the
symmetry groups for particles with rigidity and their
supersymmetric generalizations.

\bigskip
\noindent{\bf Acknowledgement}

I would like to thank the members of the Institut f\"ur
Theoretische Physik, Universit\"at Hannover, where considerable
part of this work was finished, for their hospitality. Especially
I would like to thank Prof. Olaf Lechtenfeld for his interest and
enlightening discussions. The work  was supported in part by the
Russian Foundation of Fundamental Research, under the grant
99-02-18417 and the joint grant RFFR-DFG 99-02-04022, and by a
grant of the Committee for Collaboration between Czech Republic
and JINR.
\bigskip

\section{APPENDIX.
 $N$ EXTENDED SUPERCONFORMAL ALGEBRA}
\def\theequation{A.\arabic{equation}}
\setcounter{equation}0
Having in mind the application of nonlinear realizations of $N$
extended Superconformal Algebra to the description of $N$ extended
spinning particle we describe in this Appendix the $N$ extended
SCA as subalgebra of the diffeomorphism algebra of the
$(1,N)$ superspace
$(s,\eta_a), (a=1,2,  \cdots N)$.
 The generators of the
corresponding diffeomorphism group regular at the origin can be
written in the coordinate representation as
\begin{eqnarray}                \label{rep}
{P^{(m)a_1,a_2,...,a_n}}_0\!\!&=&\!\!
is^m\eta^{a_1}\eta^{a_2}... \eta^{a_n}\frac{\partial}
{\partial s},\\\nonumber
{P^{(m)a_1,a_2,...,a_n}}_a\!\!&=&\!\!
is^m\eta^{a_1}\eta^{a_2}... \eta^{a_n}\frac{\partial}
{\partial \eta_a},  n\leq N.
\end{eqnarray}
With the help of this representation one can easily to calculate
the corresponding algebra of diffeomorphisms. All of these
generators can be naturally ordered in accordance with their
dimensionality ($dim \;s=+1,\; dim \;\eta_a=+1/2$):
\begin{eqnarray}
dim\;P_0\!\!\!&=&\!\!\!-1,\;dim\;P_a=dim\;{P^{a}}_0=-\frac{1}{2}
,\;\\\nonumber
dim\;{P^{0}}_0\!\!\!&=&\!\!\!dim\;{P^{a_1}}_a=dim\;{P^{a_1a_2}}_0=0,\quad
etc.
\end{eqnarray}

The $N$ extended SCA in the $(s,\eta_a)$ superspace is
characterized by $N$ supercovariant derivatives
\begin{equation}\label{CD}
D_a=\partial/\partial\eta^a + i\eta_a\partial/{\partial s},
\end{equation}
which transform homogeneously under the transformations of the
$N$ extended SCA. One can show, \cite{GS},
 that the corresponding infinitesimal
transformations of the coordinates $(s,\eta_a)$ can be described
in terms of an unconstrained  scalar superfunction
$\Lambda(s,\eta_a)$:
\begin{equation}\label{InfT}
\delta s=\Lambda-{1\over 2}\eta_aD_a\Lambda, \;\;\;\;
\delta\eta_a=-{i\over 2}D_a\Lambda.
\end{equation}
The composition law for two transformations with parameters
$\Lambda_1$ and $\Lambda_2$ $(\left[\delta_1,\delta_2\right]=\delta_3)$
     is
\begin{equation}\label{Composition}
\Lambda_3=\Lambda_1\partial_s\Lambda_2-\Lambda_2\partial_s\Lambda_1-
\frac{i}{2}D_a\Lambda_1D_a\Lambda_2.
\end{equation}
Each function $\Lambda$ is in one to one correspondence with some
element of the SCA. Naturally all functions $\Lambda$ are divided
on classes by their dependence on the grassmann coordinates
$\eta_a$. Their correspondence with the generators of the SCA is
illustrated by the following table
\begin{eqnarray}\nonumber
\Lambda^0_n\!\!\!\!&=&\!\!\!\!
\epsilon_n s^n \Leftrightarrow -i\epsilon_n L_{n-1},\\
\Lambda^{1/2}_n\!\!\!\!&=&\!\!\!\!
-2i\epsilon^a_n \eta_a s^n \Leftrightarrow
-i\epsilon^a_n G^a_{n-1/2},\nonumber\\
\Lambda^{1}_n\!\!\!\!&=&\!\!\!\!
2i\epsilon^{ab}_n \eta_a \eta_bs^n \Leftrightarrow
-i\epsilon^{ab}_n T^{ab}_{n},\nonumber\\
\Lambda^{3/2}_n\!\!\!\!&=&\!\!\!\!
-2\epsilon^{abc}_n \eta_a \eta_b \eta_cs^n \Leftrightarrow
-i\epsilon^{abc}_n F^{abc}_{n+1/2},\nonumber\\\nonumber
...\!\!\!\!&&\!\!\!\! ...\\\nonumber
\Lambda^{N/2}_n\!\!\!\!&=&\!\!\!\!
2(-i)^{N(N+1)/2}\epsilon^{a_1a_2\cdots a_N}_n
\eta_{a_1} \eta_{a_2}\cdots
 \eta_{a_N}s^n\nonumber\\
&&\!\!\Leftrightarrow -i\epsilon^{a_1a_2\cdots a_N}_n
R^{a_1a_2\cdots a_N}_{n-1+N/2}.\nonumber
\end{eqnarray}
All $\Lambda'$s at the left are hermitian (with hermitian
$\epsilon'$s) and normalization is chosen to get a convenient
definitions of the generators at the right.

\begin{picture}(150,90)
\unitlength=0.53mm
\put(20,50){$L_{-1}$}
\put(40,50){$L_{0}$}
\put(60,50){$\bf L_{1}$}   \put(65,53){\circle{12}}
\put(80,50){$L_{2}$}
\put(100,50){$L_{3}$}
\put(120,50){$L_{4}$}
\put(30,40){$G^a_{-1/2}$}
\put(50,40){$\bf G^a_{1/2}$}   \put(57,41){\circle{15}}
\put(70,40){$G^a_{3/2}$}
\put(90,40){$G^a_{5/2}$}
\put(110,40){$G^a_{7/2}$}
\put(40,30){$\bf T^{ab}_{0}$} 
\put(60,30){$T^{ab}_{1}$}
\put(80,30){$T^{ab}_{2}$}
\put(100,30){$T^{ab}_{3}$}
\put(120,30){$T^{ab}_{4}$}
\put(50,20){$F^{abc}_{1/2}$}
\put(70,20){$F^{abc}_{3/2}$}
\put(90,20){$F^{abc}_{5/2}$}
\put(110,20){$F^{abc}_{7/2}$}
\put(60,10){$\Lambda^{abcd}_{1}$}
\put(80,10){$\Lambda^{abcd}_{2}$}
\put(100,10){$\Lambda^{abcd}_{3}$}
\put(120,10){$\Lambda^{abcd}_{4}$}
\put(70,0){$H^{abcde}_{3/2}$}
\put(90,0){$H^{abcde}_{5/2}$}
\put(110,0){$H^{abcde}_{7/2}$}
\put(-5,50){\bf Bose}
\put(-5,30){\bf Bose}
\put(-5,10){\bf Bose}
\put(-5,40){\bf Fermi}
\put(-5,20){\bf Fermi}
\put(-5,0){\bf Fermi}
\end{picture}

\vspace{.5cm}
\hspace*{10mm}{\bf Picture 1.} {\small
Encircled are generators whose $\Omega$-form components are used
to construct the actions for $N=0,1 - (L_1)$ and $N=2 -
(G^a_{1/2})$  spinning particles
in the superfield approach.}\vspace{.5cm}\\

The first lines of the table and  Picture 1. contain
 the generators $L_n$ of the Virasoro
algebra  with $n\geq -1$. The second line contains $N$ series
$G^a_{r},\quad (r\geq -1/2$, each starting from the supercharge $G^a_{-1/2}$.
The next line starts from the generators $T^{ab}_{0}$ of the
$SO(N)$ algebra and corresponds to its Kac -- Moody
generalization. The generators
$G^a_{r}$ form the vector representation of this algebra.  All generators and
parameters with even (odd) number of indices $a$ are bosonic
(fermionic).

The algebra of some lower generators $L_n,\; G^a_r,\; T^{ab}_m,\;
 F^{abc}_s,\;
\Lambda^{abcd}_k,\; H^{abcde}_t,  \cdots $, \\
($m,n,k$ -- integer,
$r,s,t$ -- halfinteger) is as follows:
\begin{eqnarray}\nonumber
\left[L_m,L_n\right]\!\!\!\!&=&\!\!\!\!-i(m-n)L_{m+n}, \\ \nonumber
\left[L_m,G^a_r\right]\!\!\!\!&=&\!\!\!\!-i(\frac{m}{2}-r)G^a_{m+r},\\
\left[L_m,T^{ab}_n\right]\!\!\!\!&=&\!\!\!\!imT^{ab}_{m+n}, \\\nonumber
\left[L_m,F^{abc}_s\right]\!\!\!\!&=&\!\!\!\!i(\frac{m}{2}+s)F^{abc}_{m+s},\\
\nonumber
\{G^a_r, G^b_s\}\!\!\!\!&=&\!\!\!\!2\delta^{ab}L_{r+s}+(r-s)T^{ab}_{r+s},\\
\nonumber
\left[T^{ab}_m,G^{c}_r\right]\!\!\!\!&=&\!\!\!\!
                       imF^{abc}_{m+r}+i\delta^{ac}G^b_{m+r}-
i\delta^{bc}G^a_{m+r},\\    \nonumber
\{F^{abc}_r, G^d_s\}\!\!\!\!&=&\!\!\!\!
         -2\delta^{ad}T^{bc}_{r+s}-2\delta^{bd}T^{ca}_{r+s}-\\    \nonumber
\!\!\!\!&&\!\!\!\!2\delta^{cd}T^{ab}_{r+s}-2\delta^{cd}T^{ab}_{r+s}-
\Lambda^{abcd}_{r+s},\\    \nonumber
\left[T^{ab}_m,T^{cd}_n\right]\!\!\!\!&=&\!\!\!\!-i\delta^{ac}T^{db}_{m+n}+\\
\nonumber
\!\!\!\!&&\!\!\!\!\!\!i\delta^{bc}T^{da}_{m+n}+i\delta^{ad}T^{cb}_{m+n}-i
\delta^{db}T^{ca}_{m+n},\\
\nonumber
\left[T^{de}_n,F^{abc}_r\right]\!\!\!\!&=&\!\!\!\!\!\!i\delta^{da}F^{ebc}_{n+r}
+i\delta^{db}F^{eca}_{n+r}+i\delta^{dc}F^{eab}_{n+r} -\\
\nonumber
\!\!\!\!&&\!\!\!\!i\delta^{ea}F^{dbc}_{n+r}-i\delta^{eb}F^{dca}_{n+r}-\\
\nonumber
\!\!\!\!&&\!\!\!\!\delta^{ec}F^{dab}_{n+r}-in H^{abcde}_{n+r}.
\end{eqnarray}


\begin{thebibliography}{99}
\bibitem{M} R. Marnelius. Phys.Rev., {\bf D20} (1979) 2091
\bibitem{S} W. Siegel. Int.J.Mod.Phys., {\bf A3} (1988) 2713
\bibitem{P} A. Pashnev,
{\sl Massless and spinning particles as dynamics in one-dimensional
superdiffeomorphism groups,} Preprint JINR-E2-99-42, 1999. 10pp.,
e-Print Archive: hep-th/9902143
\bibitem{P1} A. Pashnev,
{\sl Infinite dimensional symmetries and particle models,}
Proceedings of the XIV-th Max Born Symposium, Karpach, Poland
1999, 143-150
\bibitem{P2} A.I. Pashnev, Czech.J.Phys., {\bf 50} (2000) 1335-1339
\bibitem{GT} V.D. Gershun and V.I. Tkach. JETP Lett.,
            {\bf 29} (1979) 320
\bibitem{HPPT}P. Howe, S. Penati, M. Pernici and P. Townsend.
         Phys.Lett., {\bf B215} (1988) 555
\bibitem{B} F.A. Berezin. Introduction to the analysis with
anticommuting variables. Moscow Univ., 1983
\bibitem{GS} A. Galperin, E. Sokatchev. Phys.Rev., {\bf D46} (1992) 714
\bibitem{BO} A.B. Borisov, V.I. Ogievetsky. Theor.Math.Phys., {\bf 21}
(1974) 329
\bibitem{IN}E.A. Ivanov, I. Niederle. Phys.Rev., {\bf D45} (1992) 4545
\bibitem{P_0}A. Pashnev, {\sl
Nonlinear realizations of the (su\-per)diffeomorphism groups,
geometrical objects and integral invariants in the superspace,}
Preprint JINR E2-97-122. e-Print archive hep-th/9704203
\end{thebibliography}
\end{document}